\documentclass[graybox]{svmult}

\usepackage{mathptmx}       
\usepackage{helvet}         
\usepackage{courier}        
\usepackage{type1cm}        
\usepackage{makeidx}         
\usepackage{graphicx}        
\usepackage{multicol}        
\usepackage[bottom]{footmisc}

\def\pdrv#1;#2'{{\D #1\O #2}}

\def\secspt{$\buildrel{\prime\prime}\over .$}

\def\H2{H$_2$}
\def\simlt{\lower.5ex\hbox{$\; \buildrel < \over \sim \;$}}
\def\simgt{\lower.5ex\hbox{$\; \buildrel > \over \sim \;$}}

\makeindex             

\begin{document}

\title*{Lessons from the Milky Way: the Kapteyn Universe}
\titlerunning{The Kapteyn Universe}
\author{P.C. van der Kruit}

\institute{Kapteyn Astronomical Institute, University of Groningen, P.O. Box 800, 9700AV Groningen, the Netherlands\email{vdkruit@astro.rug.nl}}

\maketitle

\abstract{Jacobus Cornelius Kapteyn (1851-1922) presented a model for the
distribution of stars in space together with a dynamical interpretation
in terms of an equilibrium between the gravitational field of the stars and 
their random motion and rotation. In the vertical direction Kapteyn's results
are substantially correct. Usually the Kapteyn Universe is described as
being flawed due to neglect of interstellar absorption. Kapteyn was led
to adopt this on the basis
of widely accepted evidence by Shapley on an absence of reddening of stars in
globular clusters. But another, equally important misconception was Kapteyn's
interpretation of the two Star Streams as manifestations of two groups of 
stars rotating around a center in opposite directions. This was supported by
the observation of very different mixes in stellar types in the two streams. 
Had Kapteyn adopted the absorption as he himself had determined it he would 
not have been able to arrive at a consistent picture.
}

\section{Introduction}

\begin{figure}[t]
\centering
\includegraphics[width=10cm]{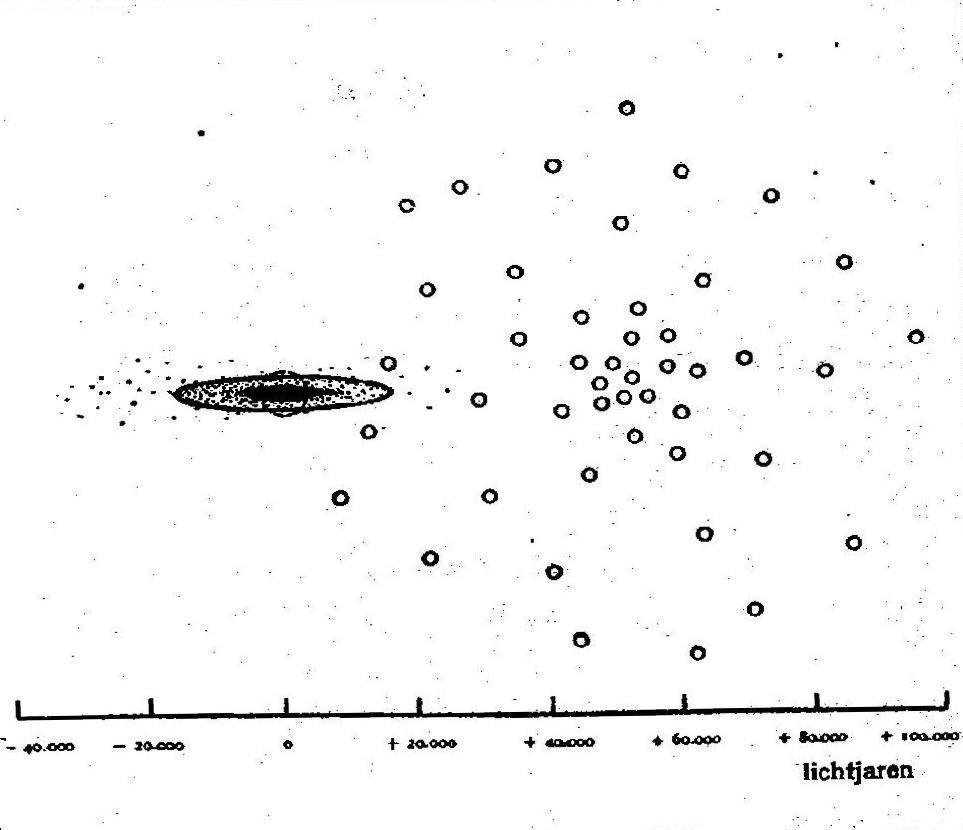}
\caption{Oort's illustration of the discrepancy of the Kapteyn Universe
and Shapley's system of globular clusters. It appeared in de Sitter;s book 
`Kosmos'.}
\label{Fig1}
\end{figure}

In 1920 Jacobus C. Kapteyn, together with his student and successor Pieter 
van Rhijn, published his well-known model for the distribution of stars in 
space, \cite{KvR20}, 
and a year later presentee a description of its 'mechanics', providing 
a consistent explanation of how the system could be in equilibrium by a 
precise balance between the gravitational force of the stars and their random 
motions and organized rotation, \cite{Kap22}. 
Ten years later the vertical dynamics was 
proved substantially correct, but in the plane of the Milky Way the 
discoveries of differential rotation and interstellar extinction had shown 
the Kapteyn Universe to be completely wrong.  Jan Hendrik Oort, who although 
completing his thesis after his death always considered himself a student of 
Kapteyn, played a major role in this with his discovery of Galactic rotation 
and the introduction of the 'Oort constants' in 1928, \cite{Oort28} 
and extended Kapteyn's work in the vertical direction with the 'Oort limit' 
in 1932, \cite{Oort32}. Oort obtained 
his doctorate under van Rhijn in  May 1926 on a study of stars of high 
velocity and accepted in November of that year a position as 'privaat-docent' 
in Leiden with a public lecture on {\it Non-light emitting matter in the 
Stellar System}. This was only four years after Kapteyn's demise and preceded 
his discovery of Galactic rotation by almost two years. In this lecture, of 
which I provided a translation in Appendix A of a Legacy volume on Kapteyn,
\cite{Legacy}, 
he concluded that the {\small {\sf 'least contrived'}} 
solution to the disparity between Shapley's system of globular clusters and 
the Kapteyn Universe (Fig.~\ref{Fig1}) 
was the assumption of an absorption of light 
in space. Why had this not been anticipated by Kapteyn? What would have 
happened had Kapteyn included an absorption correction in his modeling?

\section{The arrangement of stars in space}
Kapteyn had from his earliest research efforts had an interest in the spatial 
distribution of stars. Distances were of prime importance to this problem. 
His inaugural lecture on the occasion of his appointment of professor in 
Groningen in 1878, had the title {\it The parallaxes of the fixed stars}. And 
early on in his career he did some remarkably accurate measurements of 
trigonometric parallaxes of stars from differential timing measurements of 
meridian passages, using the Leiden meridian circle. He had not succeeded in 
securing funds for his own observatory in Groningen in spite of positive 
support from his university; Leiden and Utrecht opposed the founding of a 
third, competing observatory in the Netherlands. The stars in Kapteyn's 
sample  were selected as probably nearby on the basis of large proper motion. 
It is amazing that he was able to measure parallaxes with this method, since 
for a distance of 10 pc the effect of the annual parallax on the meridian 
passage is --depending on declination-- a few hundredths of a second of time. 

Kapteyn devised a clever method to derive absolute declinations, free from
systematic errors due to atmospheric refraction and telescope flexure. For this
he only measured differences in zenith angles of two stars at that crossed the 
meridian at about the same time at similar zenith angles on either side of the 
zenith, azimuths and times of passages through the prime vertical. 
In 1884, he found David Gill at the Cape Observatory interested in undertaking
such a project. Being struck by the possibilities of photographic plates to
register star images, Gill decided on the production of the {\it Cape 
Photographic Durchmusterung} and lured Kapteyn in offering his services to 
measure the plates. To do this Kapteyn invented his `parallactic' method in 
which a small telescope at the focal distance from the plate could be used to 
directly read off celestial coordinates on its axes.
This experience and his advisory role in the {\it Carte du Ciel}, caused 
Kapteyn to undertake  extensive photographic work on parallaxes and proper 
motions, using plates taken for him by Anders Donner of Helsinborg. 

He quickly realized that stars were too distant for direct parallax 
measurements on a grand scale and resolved to use the motion of the Sun with 
respect to nearby stars (which after all is about four radii of the Earth's 
orbit per year). To determine the arrangement of stars in space 
from star counts as a function of apparent magnitude and proper motion, he 
developed with the help of his mathematics brother Willem the necessary 
mathematical techniques, making three assumptions. (1) The 'luminosity curve', 
the distribution of absolute magnitudes, was the same everywhere, (2) there 
was no preferred direction among the motions of the stars in space, and (3) 
there was no absorption of starlight. 

\section{Star Streams}
However, around 1902 Kapteyn came to the realization that stars did not move 
completely randomly, but that there were two preferred directions
in space, corresponding to 
positions, {\small {\sf 'vertices'}}, about $100^{\circ}$ apart on the sky. 
When these were corrected for the solar velocity, they were  found to be 
roughly opposite and in the Milky Way (and in agreement with our current 
understanding, roughly in the direction of the Galactic center and anti-center).
To Kapteyn this suggested the existence of two Star Streams. He first announced 
this in 1904 at a large astronomical congress during the Louisiana Purchase 
Exhibition in St. Louis, 
and at the meeting of the British Association in Cape Town the next 
year. It was almost instantly accepted, undoubtedly helped by quick 
confirmation of the results by other astronomers, notably a young Arthur 
Stanley Eddington, who had just graduated from Cambridge. The two streams 
moved at a relative velocity, determined from radial velocities, of order 
40 km/s.

Considering the evidence and data available, this discovery is one of great
skill. And observationally it still stands. However,
it did not take long before Karl Schwarzschild came up with what we now to be 
the correct explanation, namely that of an an-isotropic velocity distribution, 
the directions of the Star Streams corresponding to the long axis of the 
velocity ellipsoid. Kapteyn was of course aware of this possible explanation, 
but all his life he believed in his own interpretation of two opposite 
streams. And he had good reasons for this. Throughout his studies the 
data available strongly suggested that the composition of the two 
streams was very different in terms of spectral types of the stars they 
contained. And Kapteyn was in good company; at the time he made his 
{\small {\sf 'first attempt'}} to determine the distribution of stars in space, 
Eddington also favored Kapteyn's interpretation over Schwarzschild's.

\section{Selected Areas}
What was needed was large samples of stars with well-determined magnitudes 
and proper motions. This could in principle be turned into mean parallaxes 
as a function of apparent magnitude and proper motions. Together these could 
then be used to find the apex of the 
solar motion and the vertices of the streams. 
Radial velocities were necessary to find the magnitude of the Sun's velocity 
in space and spectra the types of stars involved for a determination of the 
mix of ages. Kapteyn interpreted the Main Sequence as an age sequence, 
blue stars being young. Trigonometric parallaxes 
were most welcome. Realizing that one needed to go to faint magnitudes, 
Kapteyn proposed an concerted effort by many observatories by 
restricting to small areas (size depending on star surface density on the 
sky), numbering 206 distributed regularly over the sky, for which of as many 
stars as possible as many parameters as possible would be observed. It is 
tribute to Kapteyn's diplomatic skills and international standing that 
he succeeded in bringing this international coordination about.

Edward Pickering at Harvard executed a Durchmusterung of Selected Areas across 
the whole sky. He agreed to this only after Kapteyn had extended the Plan to 
include next to the `Systematic Plan' a `Special Plan' proposed by Pickering 
that concentrated on areas in the Milky Way.
Plates taken were sent to Groningen to provide positions and magnitudes. This 
work resulted in the Harvard-Groningen Durchmusterungs that provided positions 
and magnitudes down to m=16 or so in all Selected Areas. This was published 
between 1918 and 1924, and was at least partly
available for the Kapteyn \&\ van Rhijn 
analysis. But that was not deep enough. Crucial to Kapteyn's program 
was the adoption by George Ellery Hale of the Plan as the prime observational 
program for his new 60-inch telescope at Mount Wilson. From 1908 onwards 
plates were taken for all northern Selected Areas (numbers 1 through 139), 
the result of 
which only appeared in print in 1930. Even preliminary work with these data 
could only be performed after Kapteyn's death. Most of the work was done by 
Frederick Seares or under his direction. Walter Adams used the 60-inch for 
spectroscopic work, in particular to determine radial velocities. 

Hale's adoption of the Plan of Selected Areas for the 
60-inch required in his view the personal involvement of Kapteyn and the 
latter was in 1908, when the telescope became operational, appointed as 
research associate of the Carnegie Institution. Kapteyn then started to make 
annual visits to Mount Wilson until World War I prevented him 
from returning after 1914. He, however, remained research associate until his 
death.

\section{The Kapteyn Universe}
\begin{figure}[t]
\centering
\includegraphics[width=9cm]{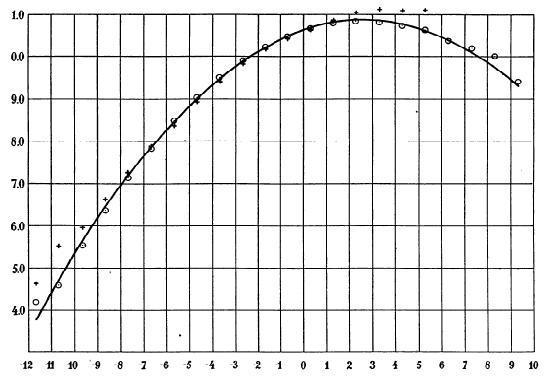}
\caption{The `luminosity curve' derived bi Kapteyn and van Rhijn in 
\cite{KvR20}. Because Kapteyn had grudgingly accepted the term parsec, he felt
obliged to redefine absolute magnitudes to a distance of 1 pc.}
\label{Fig3}
\end{figure}

The analysis in Kapteyn \&\ van Rhijn (1920), \cite{KvR20},
was preliminary. {\small {\sf 'Now 
that, after so many years of preparation, our data seem at last to be 
sufficient for the purpose, we have been unable to restrain our curiosity and 
have resolved to carry through completely a small part of the work ...'}}. From 
their data, and average parallaxes of stars as a function of magnitude
and proper  motion determined by van Rhijn, \cite{vanR20}, 
they constructed a {\small {\sf 'luminosity curve', see Fig.~\ref{Fig3}}}, 
which was assumed to be the same everywhere in space. 

Using mathematical methods partly developed by Karl 
Schwarzschild, they solved the counts by transforming these into a density of 
stars as a function of distance. This process was done for four ranges of 
latitude, $0^{\circ}$, $30^{\circ}$, $60^{\circ}$ and $90^{\circ}$. The 
(preliminary) solution only fitted to densities as a function of distance from 
the Sun, so effectively assumed that the Sun was at or close to the 
symmetry point of the system. The result found by Kapteyn and van Rhijn is in 
the top picture of Fig.~\ref{Fig2}.

\begin{figure}[t]
\centering
\includegraphics[width=0.98\textwidth]{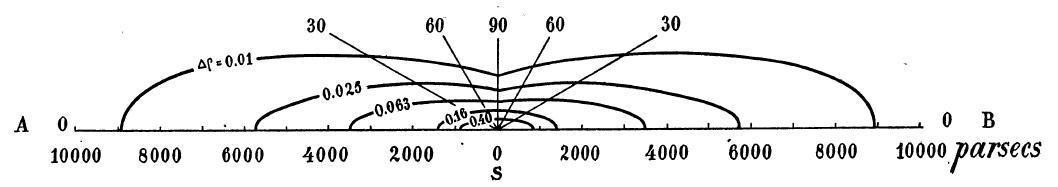}
\includegraphics[width=0.98\textwidth]{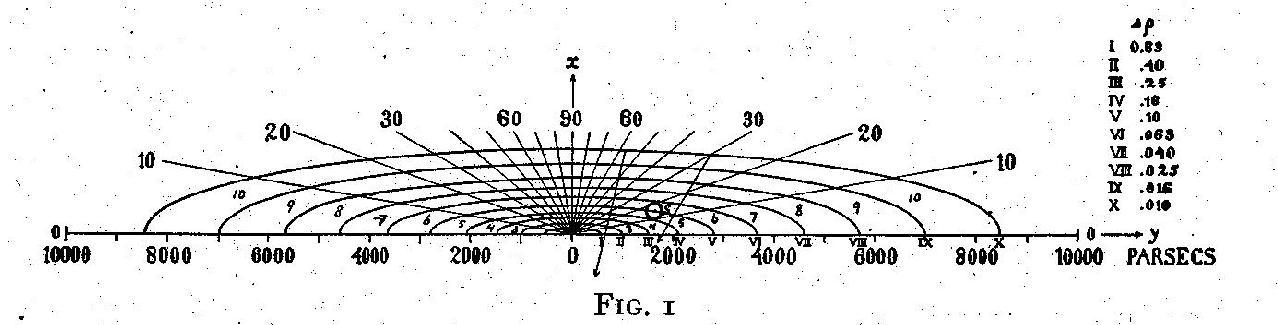}
\caption{On top the distributions of stars in space as determined by
Kapteyn and van Rhijn (1920). The solution involves a dependence only with 
distance from the Sun at latitudes $0^{\circ}$, $30^{\circ}$, $60^{\circ}$ and 
$90^{\circ}$. The bottom illustration is from Kapteyn (1922), where he fitted
ellipsoids to the densities to facilitate computation of the gravitational 
potential. His interpretation of the Star Streams as two opposite rotations led
him to adopt a position of the Sun 650 pa away from the center and on the basis
of Hertzsprung study of the distribution of Cepheids on the sky at 38 pc from
the plane. The Sun is at the circle designated `S'.}
\label{Fig4}
\end{figure}

\begin{figure}[t]
\centering
\includegraphics[height=7cm]{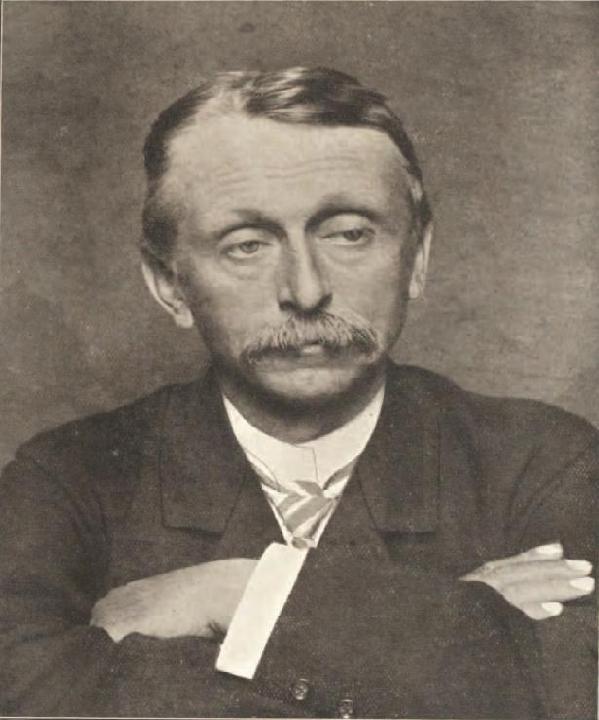}
\includegraphics[height=7cm]{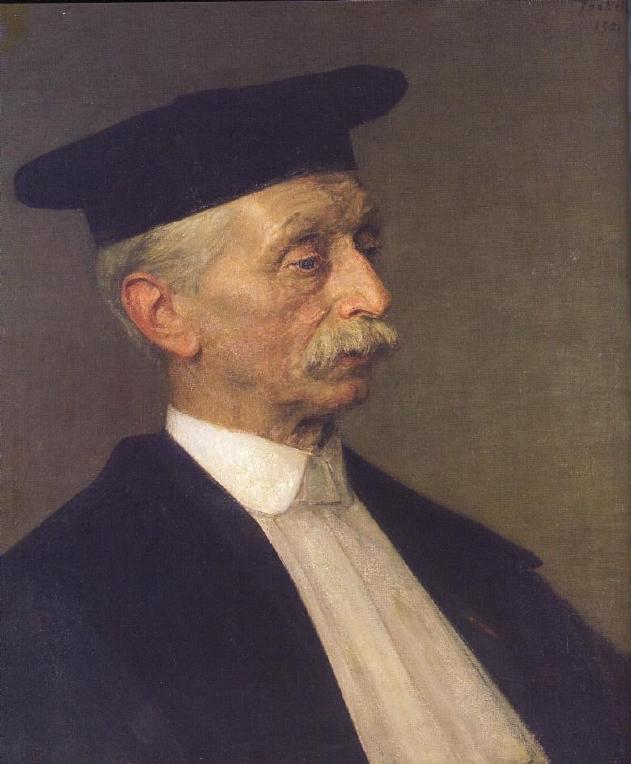}
\caption{Jacobus Cornelius Kapteyn (1851--1922) at age 40 and 70.}
\label{Fig2}
\end{figure}

With this result Kapteyn could do what he long had been preparing for and that 
was to study the {\small {\sf `mechanics'.}} He assumed that the system was 
in equilibrium. In his 1922 paper, \cite{Kap22}, 
he first assumed that the equi-density
surfaces were ellipsoids (see Fig.~\ref{Fig4}, bottom). He did this so that
he could calculate relatively straightforwardly the potential assuming that 
the star distribution and that of the density of gravitating matter was the 
same. He used a vertical mean velocity of stars of 10 km/sec (which 
corresponds to a velocity dispersion of 12 km/sec). In the vertical direction 
he then found that the system was in equilibrium if the mass of an average
star was between 1.4 and 2.2 solar masses. The
average measured mass of a binary was 1.6 in these units, so that if all stars 
were binaries there would be no need to invoke {\small {\sf `dark matter in the
Universe'}}. At least, {\small {\sf `this mass cannot be excessive'}}. 
This is a first
application of what we now call stellar or galactic dynamics. In modern terms, 
using the isothermal sheet description of van der Kruit \&\ Searle, \cite{vdKS},
Kapteyn's density distribution is fitted pretty well with $z_{\circ}$=650 pc, 
rather close to the 700 pc or so we currently use. His velocity dispersion is
a bit low compared to the modern 17 km/sec or so.

For the horizontal direction, equilibrium could only be achieved if there
were some rotation. To Kapteyn it seemed natural that this was the phenomenon
of his Star Streams; about half the stars rotating 
in one direction and the other in the opposite one. The direction towards the 
(rotation) center would be perpendicular to the streams and could then
be in only two 
directions, one of which was Carina and indeed there the Milky Way is very 
bright. The relative velocity was 40 km/sec, so it seemed natural 
to assume that the rotation velocity was 20 km/sec. Indeed, when he estimated
what it had to be with the mass densities that he derived from the vertical
equilibrium he found 18-20 km/sec! But for this to work, the Sun had
to be away from the center. Kapteyn argued that this implied a distance up to
1.5 kpc, but for reasons having to do with details of the solutions in
the density distributions he settled for 
650 pc. For the vertical displacement he assumed 38 pc, based upon a study of 
the distribution of Cepheids by his son-in-law Ejnar Hertzsprung. 
The resulting position of the Sun has been indicated
with the `S' near the small circle in Fig.~\ref{Fig4}, bottom.

So we see, that the adoption of this model by Kapteyn had also a dynamical
basis, partly correct (in the vertical direction), partly incorrect (due to
his interpretation of the Star Streams). But it was ingenious and his 
introduction of a dynamical framework farsighted.

\section{Kapteyn's studies of absorption}
Kapteyn had always worried about absorption. Over his career he wrote four 
papers on this subject, one in the Astronomical Journal, three in the 
Astrophysical Journal. In 1903, Pickering, \cite{Pick03},
 had found that 
the star ratio (the ratio of numbers of stars between two consecutive 
magnitudes) deviated from the `theoretical' value for a uniform distribution 
of stars in space. He suggested that there was significant absorption, amounting
in Kapteyn's notation to 0.18 magnitudes per unit distance. Since this unit is 
the distance corresponding to a parallax of 0\secspt1, this is an extremely 
large amount.
Moreover, in 1904 George Comstock found that the mean proper motion of stars 
at a fixed magnitude depended on Galactic latitude (the smallest in the Milky 
Way), which also would indicate interstellar absorption, \cite{Com04}. 
In 1904, Kapteyn, \cite{Kap04},
investigated these claims, showing how sensitively a small amount of 
extinction would change the inferred stellar distributions. He used the slope 
of the luminosity curve as a fiducial point to link to observed counts and 
concluded that some absorption indeed was likely. But he objected against the 
(extremely large) values of Comstock, since these would put the Sun in a very 
special place, namely near an extreme minimum in the star density.

In 1909 Kapteyn wrote two papers in the Astrophysical Journal on absorption,
\cite{Kap09a}, \cite{Kap09b}. 
There was not much convincing evidence for absorption by gas (in stellar 
spectra), but he speculated that absorption by dust would result in a 
reddening of the starlight. He found values corresponding to 0.001 magnitude 
per parsec in the 
photographic band (and 0.0005 in the visual), which at present is a bit small 
but not far off the mark. In a paper in 1914, in which he returned to the 
issue by reviewing its status at the time, he reaffirmed this but concluded 
that from observational data it could not be proven that this was not
be the result of a correlation of color with absolute magnitude, \cite{Kap14}.

The deciding piece of evidence, however, was Harlow Shapley's result in 
1916, \cite{Shap16}, that the colors of the stars in the globular 
cluster M13 (at 10 kpc) should be 2.5 magnitudes redder than they were 
observed to be if Kapteyn's value for the absorption were adopted. Space was, 
according to Shapley, transparent. This must have convinced Kapteyn (and with 
him others) that absorption was only of secondary importance.

Shapley went on to map the distribution of globular clusters and found
them to be scattered over a somewhat ellipsoidal volume with the center
in the direction of Sagittarius, roughly perpendicular to Kapteyn's center,
and at a distance of 20 kpc, see \cite{Shap19} and Fig. 1 therein. Shapley's 
distance to the center was too large, as we now know, to some extent also as a
result of absorption since many clusters are at a somewhat low latitude.

What did Kapteyn himself think of Shapley's result? In one of his last papers,
written with van Rhijn in 1922, \cite{KvR22},
it was argued that Shapley's distances were
too large because the local calibrating Cepheids of short periods 
had high proper motions and 
therefore had to be near and thus fainter than Shapley's variables in the
globular clusters. The longer
period variables were giving a different result and Kapteyn and van Rhijn
argued that this result was inferior. It was supported by work of Kapteyn's
student Schouten, who derived distances to globular clusters by adopting
the same luminosity curve as locally. This of course ignored the fact that 
locally many stars are dwarfs, while all stars used in the clusters were 
giants. 

\section{What if Kapteyn had not neglected absorption?}
I have been asked this question regularly. Obviously we do not know for sure.
Kapteyn presented his model for the Sidereal System less than a year before 
his death. It was in accord with current 
knowledge and provided a consistent description. 
As a conclusion of this 
contribution I quote from the draft of my soon (later this year) 
to appear biography {\it Jacobus Cornelius
Kapteyn: Born investigator of the Heavens}, \cite{vdk14}.

Finally, one might wonder what would have happened had Kapteyn adopted
interstellar absorption. That would have been much in violation of Shapley's
result and Kapteyn would for some reason or the other have had to assume
it depended on latitude. He did find values for extinction in his earlier
work that in hindsight are not unreasonable, but he derived that from relatively
bright stars all over the sky. No latitude dependence was evident, and it
seems unlikely that he would have detected any had he looked for it.

Had Kapteyn applied an absorption correction he would have found a 
significantly flatter vertical profile of star densities, which would have 
lowered the deduced average mass of a star, destroying the dynamical 
consistency that he found. Similarly, a much flatter radial density profile 
would have made the picture of the rotation as revealed by the star streams 
more problematic. If it would have had a central point of maximum density it 
would be further away than in the Kapteyn a universe and a rotation speed of 
20 km/sec as revealed by the Star Streams would have to be linked to a 
rotation around a more distant center. As it was, Kapteyn's solution of such a 
rotation around a center a half to one kiloparsec away dynamically made sense, 
and it is not obvious that such a picture could be constructed with so 
small a rotation in an extended stellar system. [...]

So there were at least {\it two} fundamental issues involved. Kapteyn's 
Universe was not only false because his assumption of transparent space turned 
out incorrect, but also his conviction that the star streams did not result 
from an anisotropy in the random stellar motions was essential for his 
arriving at a consistent picture. The Kapteyn Universe was built on both 
pillars and both eventually had to be replaced.

Kapteyn's picture of the Sidereal System was difficult to avoid. Lodewijk
Woltjer in \cite{Legacy} referred to it as 
{\small {\sf 'Kapteyn's unfortunate Universe'}}, on account of 
{\small {\sf `the unlucky moment at 
which Kapteyn presented his Universe and to the perhaps somewhat unphysical 
interpretation he gave to the two `Star Streams', which had contributed so 
much to his fame'.}} It is more uncertain to speculate what would have happened
had Kapteyn lived a decade or so longer and would have been around when 
Galactic differential rotation and interstellar absorption were established. 

\section{Lessons?}

My chapter in \cite{Legacy} started as follows:
{\sf {\small `Around 1990 the Kapteyn Astronomical Institute decided to start a
preprint series and we chose to display on the covers a picture of
Kapteyn. Some of us were told privately and discretely by a few of
our colleagues abroad (mostly in the U.S.) that is was inappropriate
to use this picture. After all, Kapteyn had been proven wrong in almost
all respects; his work on interstellar absorption was 
shown to seriously underestimate the effect, while his model of the
distribution of stars in space and his
description of the kinematics and dynamics in terms of circular motions
in two star streams was also incorrect. A consensus
among our staff eventually led to the replacement of that picture by a
`more current' one; I regret having not more strongly opposed that then.'}}
It was the picture at the right in Fig.~4, and I still do regret this. 

Owen Gingerich, in the same volume, wrote:
{\sf {\small `For American college students studying introductory astronomy, 
the chances are about
even that they will encounter the name of Kapteyn.  The distinguished 
astronomer of Groningen is
mentioned in seven of ten recent textbooks that I examined, always 
associated with the Kapteyn
Universe, and invariably as a foil for Shapley's larger and 
more modern conception of the Milky Way Galaxy. While the presentation in 
American textbooks places Kapteyn and Shapley as intellectual
rivals and their systems as totally antithetical, the historical 
reality is much different.'}}
 
So, two lesson are: `Don't take historical introductions in elementary
textbooks too seriously. And be warned that your colleagues may very well 
have done that anyway'.
\bigskip

\noindent {\small {\bf Acknowledgement.} 
I am very grateful to Ken Freeman for presenting the 
review when, due to family affairs, I had to cancel my participation to the
conference at the last moment.}

%

\end{document}